\documentclass[a4paper]{jpconf}
\usepackage{graphicx}
\usepackage{lineno}
\begin{document}
\title{Some results of test beam studies of Transition Radiation Detector 
prototypes at CERN}

\author{
 V~O~Tikhomirov$^{1,2,}$\footnote[8]{To whom any correspondence should be addressed.},
 T~Brooks$^{3}$,
 M~Joos$^{3}$, 
 C~Rembser$^{3}$, 
 E~Celebi$^{4}$,
 S~Gurbuz$^{4}$,
 S~A~Cetin$^{5}$,
 S~P~Konovalov$^{2}$,
 K~Zhukov$^{2}$,
 K~A~Fillipov$^{1,2}$, 
 A~Romaniouk$^{1}$, 
 S~Yu~Smirnov$^{1}$, 
 P~E~Teterin$^{1}$, 
 K~A~Vorobev$^{1}$,
 A~S~Boldyrev$^{6}$, 
 A~Maevsky$^{6}$,
 and D~Derendarz$^{7}$}

\address{$^1$ National Research Nuclear University MEPhI (Moscow Engineering 
Physics Institute), Kashirskoe highway 31, Moscow, 115409, Russia}
\address{$^2$ P.N.~Lebedev Physical Institute of the Russian Academy of 
Sciences, Leninsky prospect 53, Moscow, 119991, Russia}
\address{$^3$ CERN, the European Organization for Nuclear Research, 
  CH-1211 Geneva 23, Switzerland}
\address{$^4$ Bo\u{g}azi\c{c}i University,34342 Bebek/Istanbul Turkey}
\address{$^5$ Istanbul Bilgi University, High Energy Physics Research Center, 
  Eyup, Istanbul, 34060, Turkey}
\address{$^{6}$ Skobeltsyn Institute of Nuclear Physics Lomonosov Moscow State 
  University, Moscow, Russia}
\address{$^{7}$ Institute of Nuclear Physics Polish Academy of Sciences, 
  Krakow, Poland}

\ead{Vladimir.Tikhomirov@cern.ch}

%------------------------------------------------------------
\begin{abstract}

Operating conditions and challenging demands of present and future accelerator 
experiments result in new requirements on detector systems. 
There are many ongoing activities aimed to develop  new technologies 
and to improve the properties of detectors based on existing technologies. 
Our work is dedicated to  development of Transition Radiation Detectors 
(TRD) suitable for different applications.
In this paper results obtained in beam tests at SPS accelerator at CERN with 
the TRD prototype based on straw technology are presented. 
TRD performance was studied as a function of thickness of the transition 
radiation radiator and working gas mixture pressure.
\end{abstract}

%------------------------------------------------------------
\section{Introduction}

\begin{figure}[ht]
\begin{minipage}{18pc}
\includegraphics[width=18pc,height=15pc]{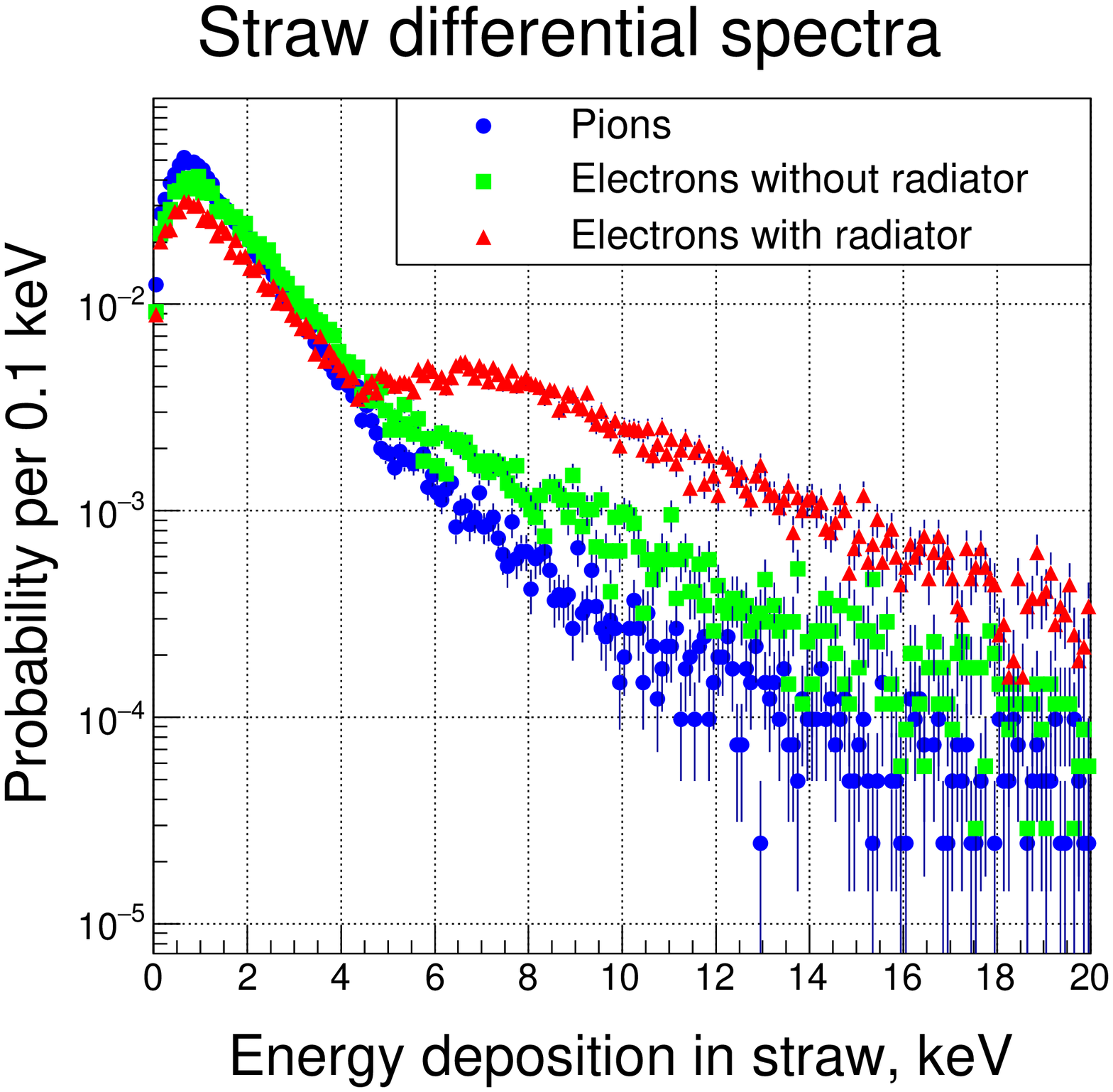}
\caption{\label{fig:diff_spectra}Examples of energy spectra in single straw
chamber. Incident beam particles are 20~GeV pions or electrons.}
\end{minipage}\hspace{2pc}
\begin{minipage}{18pc}
\includegraphics[width=18pc,height=15pc]{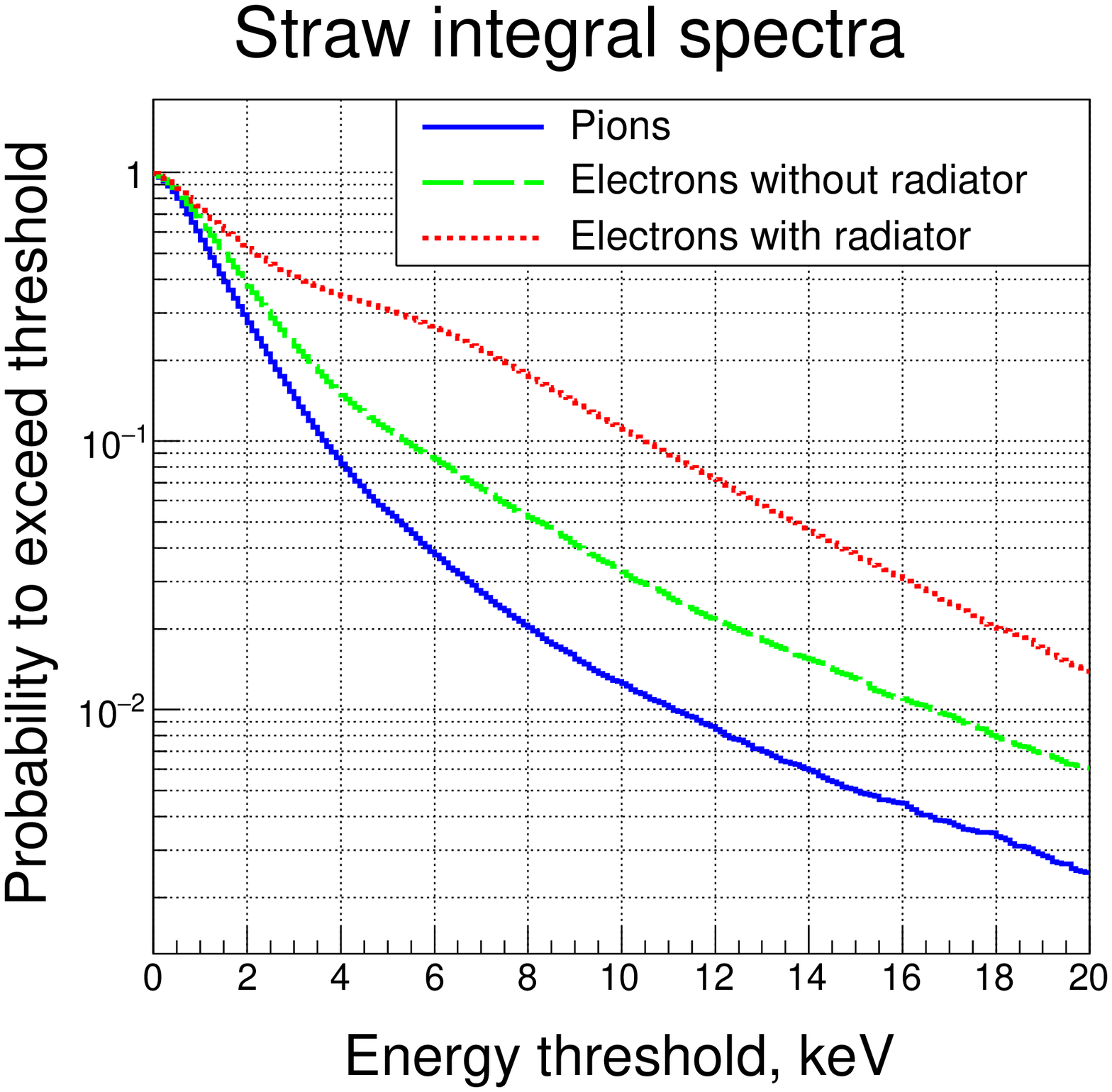}
\caption{\label{fig:integ_spectra}Integral energy spectra --
probability to exceed some energy deposition in a single straw.}
\end{minipage} 
\end{figure}

\begin{figure}[ht]
\includegraphics[width=18pc,height=15pc]{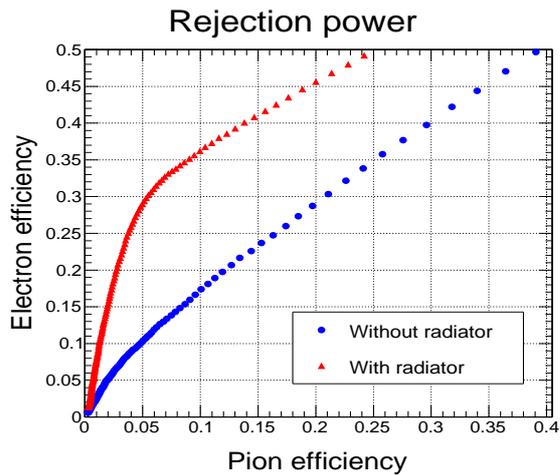}\hspace{2pc}
\begin{minipage}[b]{18pc}\caption{\label{fig:el_vs_pi}Efficiency of electron identification vs efficiency of pion identification.}
\end{minipage}
\end{figure}

Particle identification with  TRDs  is based on the difference of 
energies deposited in detector module(s) crossed by particles  with 
different gamma factors.
Particles with high Lorentz factor, like electrons, 
produce  photons in Transition Radiation (TR) radiator which are absorbed in 
the detector sensitive volume.
For detectors which have small thickness like straw based detector 
(straw tube diameter of 4 mm)  averaged energy deposition 
due to ionization losses of particle is significantly less than energy of 
TR photon which is above $\sim$5~keV. With some probability this allows 
to separate events with TR from events where TR was not absorbed.
Typical resulting spectra of energy depositions for different particle types 
in straw detector are shown in figure~\ref{fig:diff_spectra}. 
Figure~\ref{fig:integ_spectra} shows the same spectra in
integral form: probability to exceed some energy threshold as a function 
of threshold. One sees large difference
for pions and electrons when TR radiator is used. Very often for the 
comparison of performances of different detectors with 
the same structure it is better to use presentation shown 
in figure~\ref{fig:el_vs_pi}. 
Each point on this plot correspond to some threshold and projections on 
axis show probability for electron and pion to exceed this threshold. 
The larger electron efficiency at the same pion efficiency the better 
performance of the detector (better electron/pion separation). 
Optimal operation point of the detector (best separation) is around 
a ``knee'' of the dependence for electrons. In our case this point  
is around 0.05 of pion efficiency. 
See review in \cite{PDG} for more details on principles and usage of TRDs. 

 In order to improve particle identification, several parameters 
of the TR radiator and detector system can  be tuned depending on 
the desired physics task. 
The most important ones are radiators parameters, active gas composition 
and its thickness. 
One should also mention that the method of data analysis plays significant 
role in final particle separation process. 
This paper presents the results on the studies of detector performance 
with different radiator thicknesses  
and different working gas pressures.
Some other results obtained with the TRD prototype can be also found in 
reports \cite{Celebi} and \cite{Tishchenko}, presented at this Conference.

%------------------------------------------------------------
\section{Test beam set-up}

\begin{figure}[ht]
\includegraphics[width=27pc]{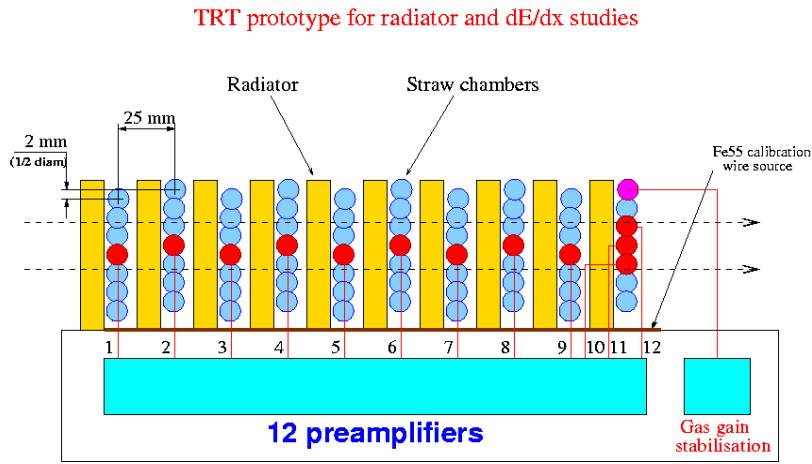}\hspace{2pc}
\begin{minipage}[b]{9pc}\caption{\label{fig:strawTRDsetup}Schematic view of the TRD prototype exposed to the test beam.}
\end{minipage}
\end{figure}

Schematic view of the test beam straw TRD prototype is shown in 
figure~\ref{fig:strawTRDsetup}. Beam particle, triggered by  
10$\times$10~mm$^2$ scintillators, crosses ten straw layers. 
The gaps between layers are used to install different TR radiator blocks. 
 Thin-walled proportional chambers 
(straws) are used to measure ionization and TR photons spectra.
Straws with 4~mm diameter are made from a special conductive Kapton film. 
Straw wall has thickness of 70~$\mu$m. 
Similar chambers are used in the Transition Radiation Tracker 
detector~\cite{TRT} of the ATLAS experiment~\cite{ATLAS} at LHC.
The straws in the prototype were  operated with a gas mixture of 
71.8\% Xe, 25.6\% CO$_2$ and 2.6\% O$_2$.
The gas gain was of about 2.5$\cdot10^4$ and it was controlled with 
an accuracy of about 1.5\% during the run using Fe$^{55}$ 
source.   Signals from straws were recorded using VME QDC modules. 
In order to separate signals from noise only energy depositions above  
 100~eV were considered.

Each radiator block contains 36 polypropylene foils of 15~$\mu$m
thick spaced by 213~$\mu$m air gap.  Total radiator thickness
along the beam direction is 8.2~mm. Two different radiator configurations were
considered. In the first case each of 10  radiators were
situated in front of each of 10 straw layers. For the second setup 
(``double radiators'') 10 radiators were grouped by two and installed in
front of last five straw layers.

One should note that  TR spectrum is changed with an increase of a number 
radiator-detector layers
until it reaches saturation which is defined by equilibrium between
produced and absorbed TR photons. That is why all results presented here are 
related to TR spectra obtained in the last straw layer where the saturation 
is guaranteed. Ionization spectra of pions are not affected by TR and 
therefore in order to increase statistics pion 
spectra from all 10 layer of straws were merged.

%------------------------------------------------------------
\section{Results}
\label{sec:Results}

\begin{figure}[ht]
\includegraphics[width=20pc,height=18pc]{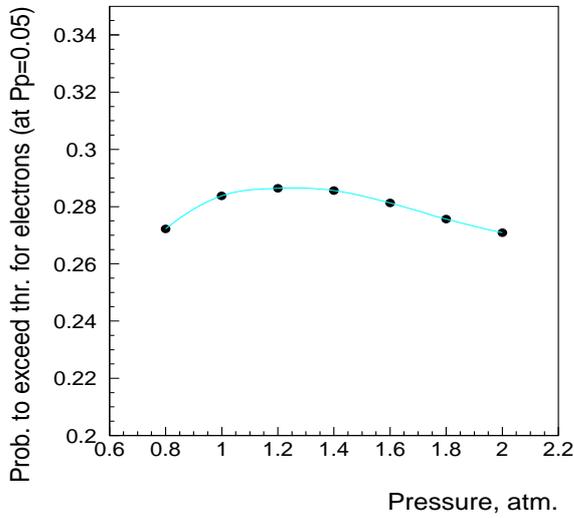}\hspace{2pc}
\begin{minipage}[b]{16pc}\caption{\label{fig:MCpressure}Monte Carlo simulation
of expected electron registered efficiency at 0.05 pion efficiency as
a function of working gas pressure. 
Line connecting points is guide to the eyes only.}
\end{minipage}
\end{figure}

As it was mention above  energetic TR photons
may  cross  straw layers without absorption and can be absorbed 
by the following radiator blocks. In order to increase a registration 
efficiency of the
energetic photons one can increase the pressure of the working gas
in the straws. It will increase a probability  of TR  absorption 
in the detector. 
However this would also increase  ionization losses of particles.
That is why the resulting effect of the pressure increase on the particle 
separation is not evident. Preliminary Monte
Carlo simulations predict rather week dependence of probability for 
electron to exceed registration 
threshold at fixed probability for pions -- figure~\ref{fig:MCpressure} 
with maximum close to 1.2 bar (absolute pressure). 
A special test with the gas pressure of 1.5 bar (abs.) was carried out 
to  verify  this behavior.

\begin{figure}[hb]
\begin{minipage}{18pc}
\includegraphics[width=18pc,height=16pc]{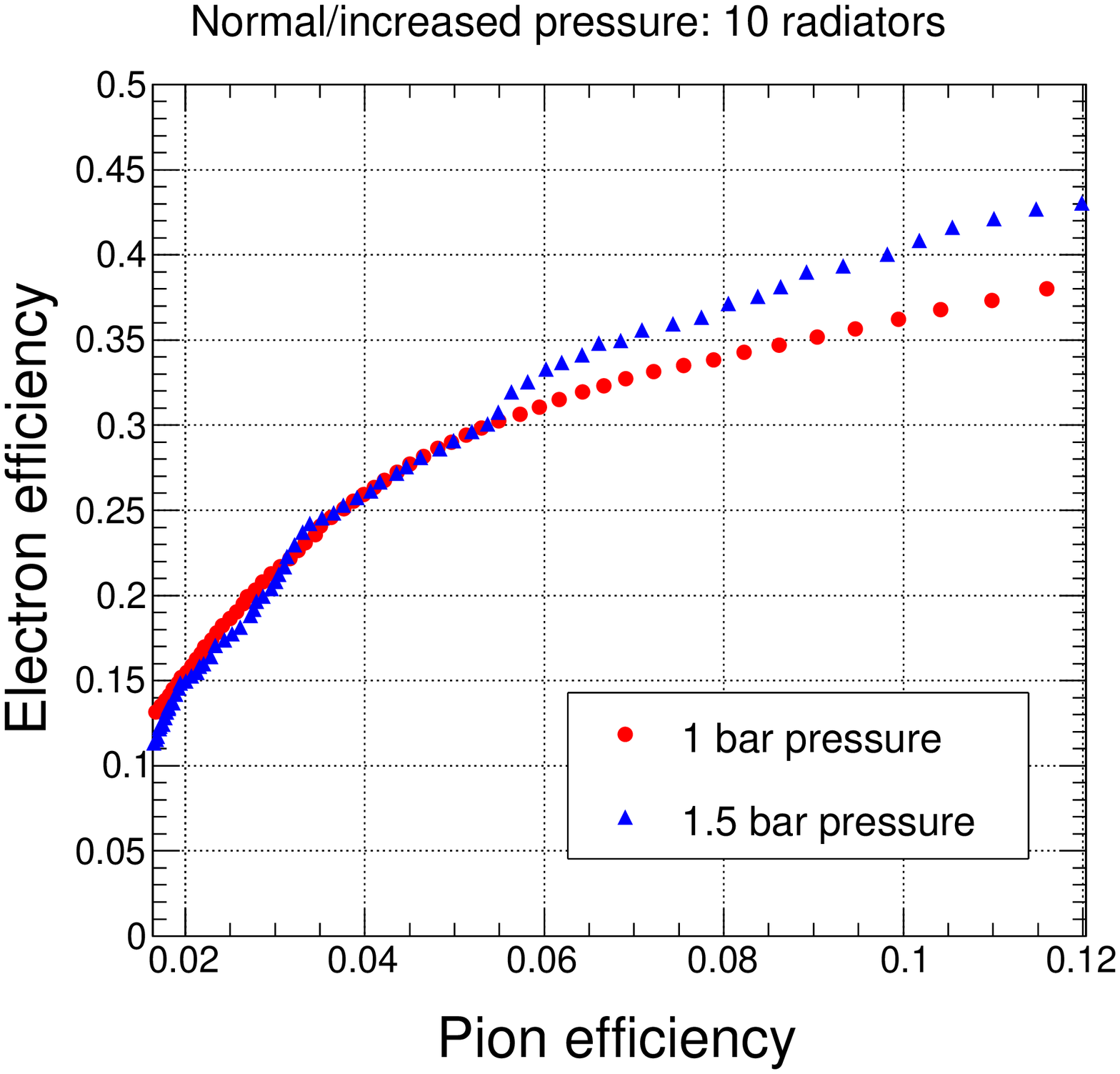}
\caption{\label{fig:pressure10}Comparison of electron vs pion efficiency at
normal and increased gas pressure in straws. Setup with 10 single radiators.}
\end{minipage}\hspace{2pc}
\begin{minipage}{18pc}
\includegraphics[width=18pc,height=16pc]{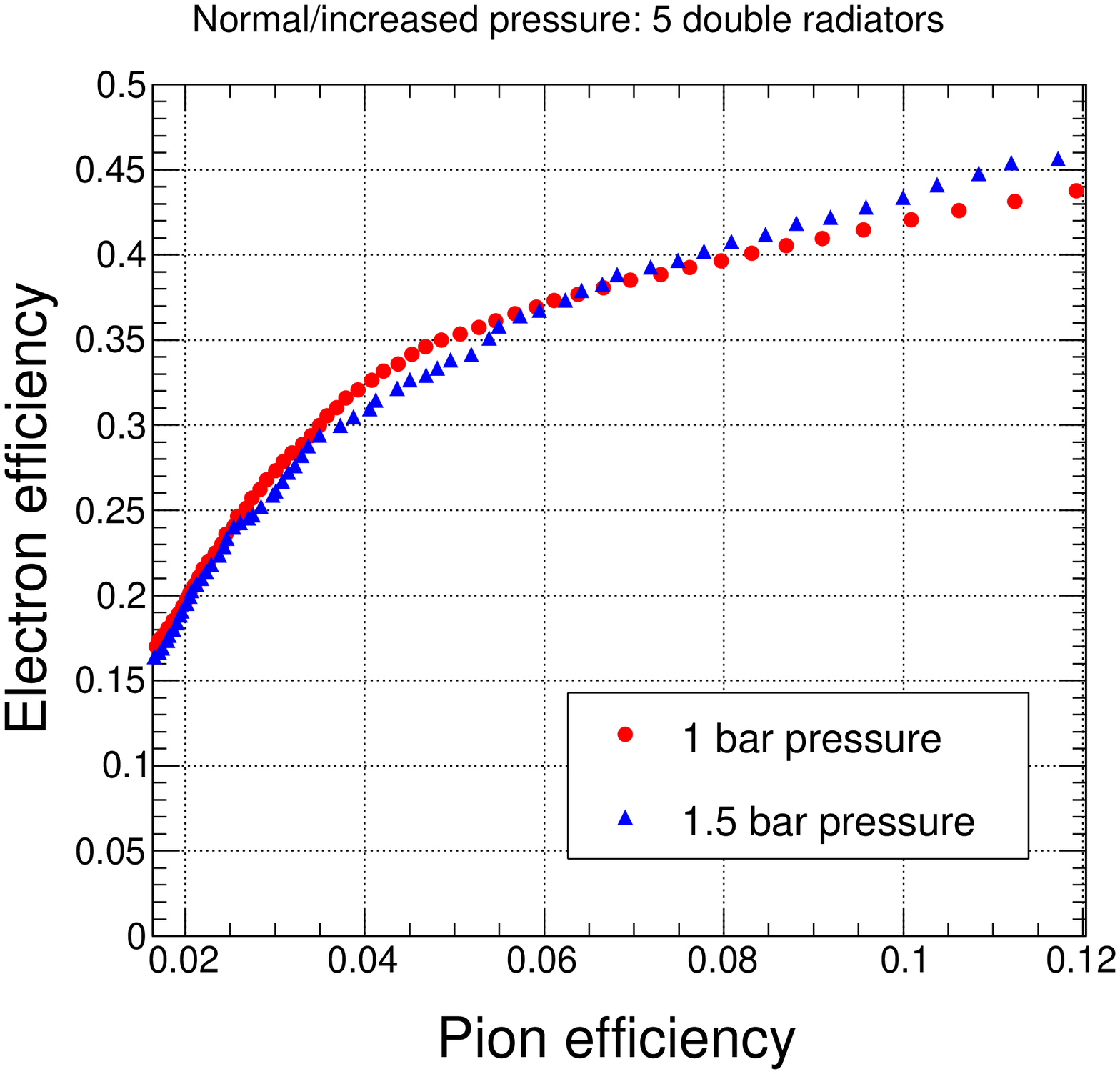}
\caption{\label{fig:pressure5}Comparison of electron vs pion efficiency at
normal and increased gas pressure in straws. Setup with 5 double radiators.}
\end{minipage} 
\end{figure}

Figure~\ref{fig:pressure10} shows comparison of probability for electrons 
vs probability for pions
 to exceed certain threshold at 1 bar (abs.) and 1.5 bar (abs.) of 
working gas pressure for setup  with 10 single radiator blocks.
The same comparison for setup with five double radiator blocks is presented on
figure~\ref{fig:pressure5}. One sees that in both cases the increase of
gas pressure does not significantly change these dependencies even at 
rather high photon energies which correspond
to low pion probabilities and this behavior is what was expected from 
MC simulations.

\begin{figure}[ht]
\begin{minipage}{18pc}
\includegraphics[width=18pc,height=16pc]{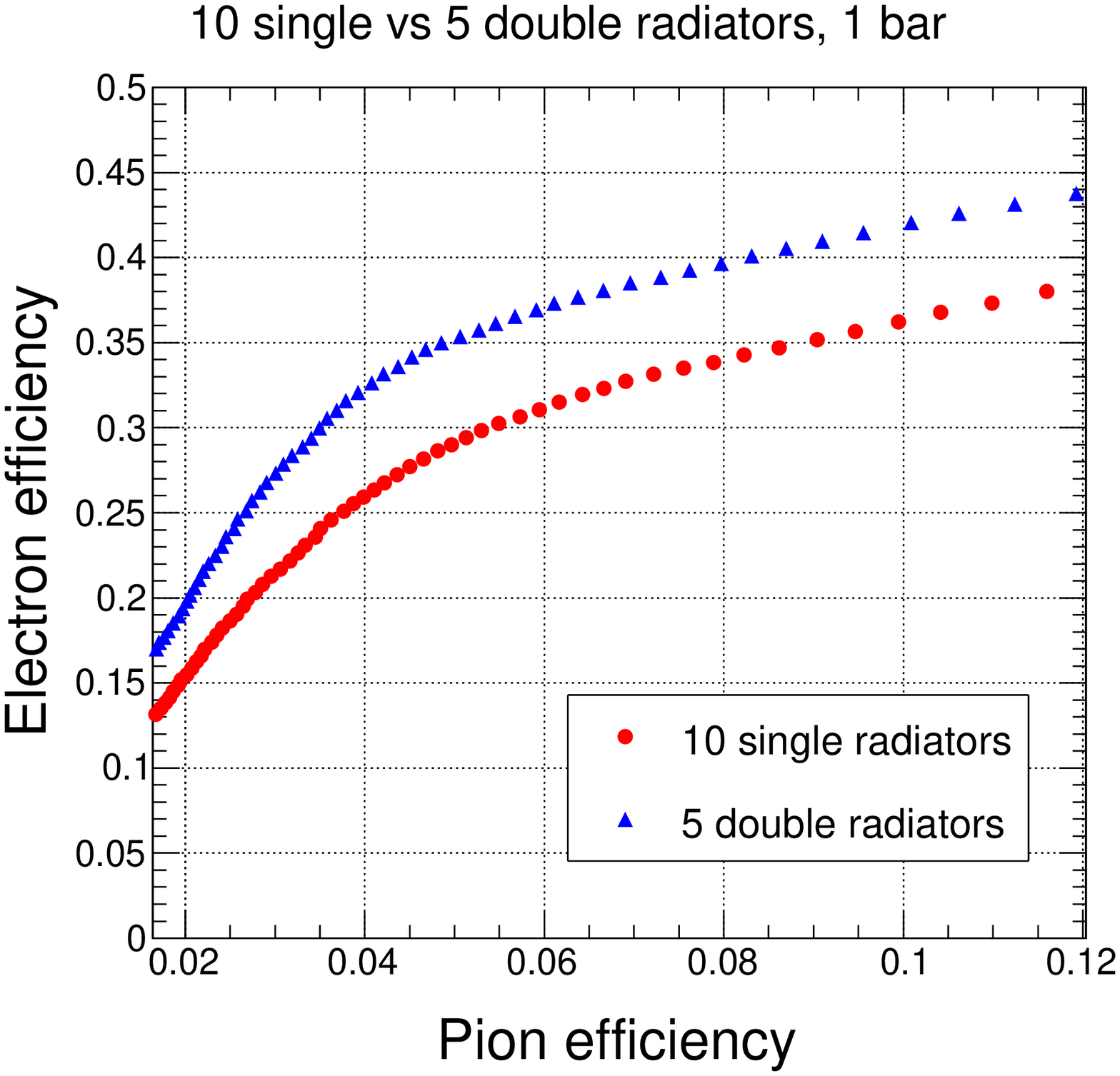}
\caption{\label{fig:5and10norm_pressure}Comparison of electron vs pion 
efficiency for 10 single and 5 double radiator setups at normal gas pressure 
in straws.}
\end{minipage}\hspace{2pc}
\begin{minipage}{18pc}
\includegraphics[width=18pc,height=16pc]{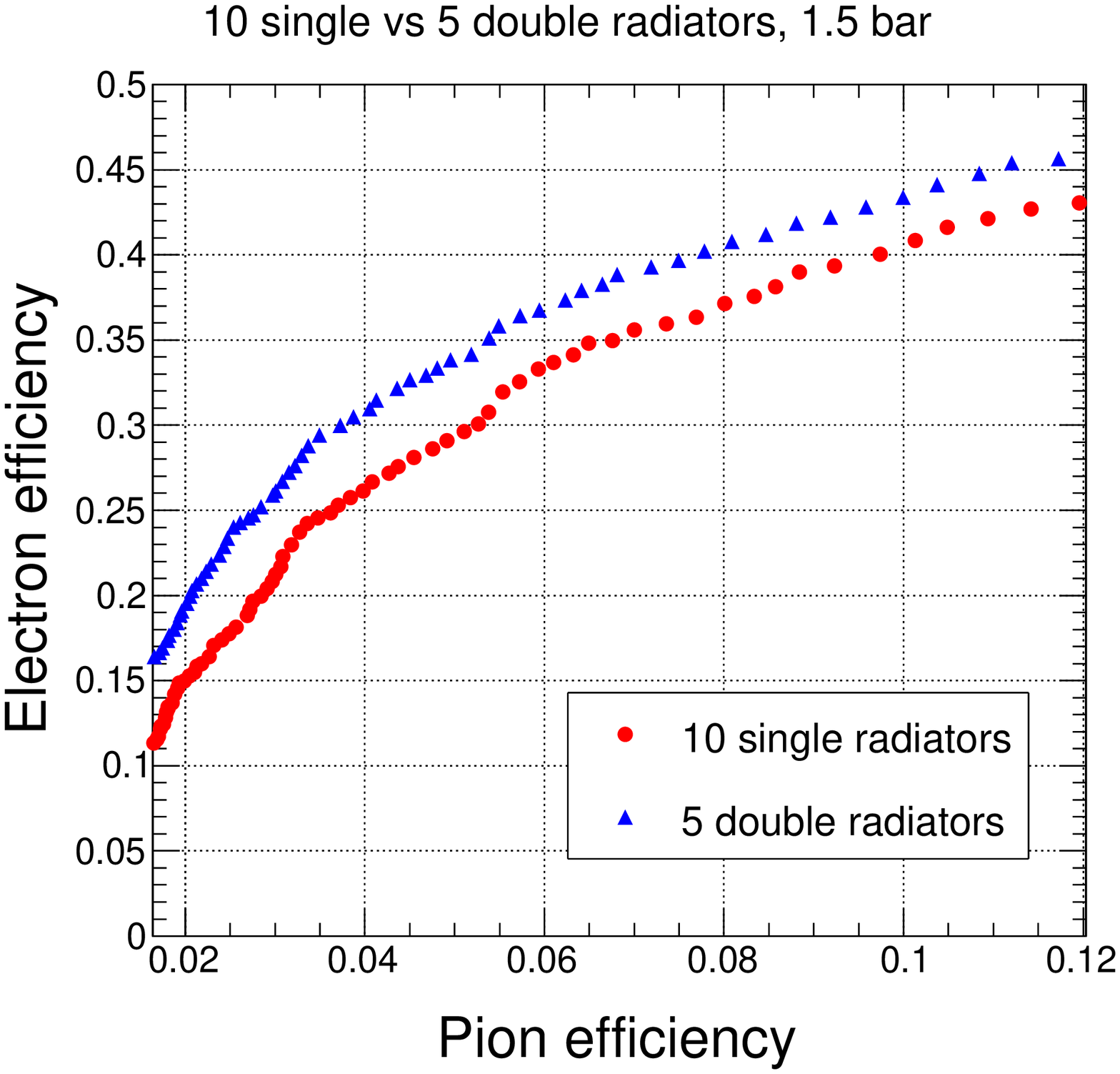}
\caption{\label{fig:5and10enlarge_pressure}Comparison of electron vs pion 
efficiency for 10 single and 5 double radiator setups at increased gas pressure 
in straws.}
\end{minipage} 
\end{figure}

Another way to improve particle identification power is to optimize the
radiator thickness. The thicker radiator the more TR photons generated but 
in that case at the same number of detector modules the total detector 
size and material budget is increased.
Figures~\ref{fig:5and10norm_pressure} and \ref{fig:5and10enlarge_pressure}
give comparison of electron vs pion efficiency dependencies for single
and double layer radiator -- at 1 and 1.5 bars of the gas pressure respectively.
In both cases thicker radiator can provide significantly better identification 
power but the length of the detector is increased by factor of 1.7.
If the total detector length is fixed then the thicker radiator will produce 
higher number of TR photons, however the reduced number of sensitive layers 
will lead to larger statistical fluctuations of the registered signal thus 
decreasing the identification performance.  
In order to reach better particle identification performance the optimal 
configuration of radiator thickness and detector sensitive volumes can be 
obtained taking into consideration all external requirements such as 
available space and total amount of material which is crossed by particle.

%------------------------------------------------------------
\section{Conclusion}
\label{sec:Conclusion}

Transition Radiation Detector prototype based on straw tubes was 
tested with 20 GeV pion and electron beams at CERN SPS accelerator. 
It was shown that an increase of the working gas pressure by  0.5 bar 
has no advantage 
for electron/pion separation  with respect to normal pressure conditions.
Better rejection power can be obtained by increasing radiator thickness
due to larger number of generated TR photons at the cost of the increased
detector length. 

%------------------------------------------------------------
\ack
%\subsection{Acknowledgments}
We gratefully acknowledge the financial support
from Russian Science Foundation -- grant No.16-12-10277. 

%------------------------------------------------------------


\section*{References}
\begin{thebibliography}{9}

\bibitem{PDG} C Patrignani {\it et al.} (Particle Data Group) 2016
Chin. Phys. C, 40, 100001, chapter ``Transition radiation detectors''.

\bibitem{Celebi} E Celebi {\it et al.} 
Test beam studies of the TRD prototype filled with different gas mixtures 
based on Xe, Kr, and Ar.
\textit{J. Phys.: Conf. Series} (In this Proceedings)

\bibitem{Tishchenko} A Tishchenko {\it et al.} 
Effect of graphen monolayer on the transition radiation yield of the radiators 
based on polyethylene foils.
This Conference, http://indico.cfr.mephi.ru/event/4/session/18/contribution/317

\bibitem{TRT} E Abat {\it et al.} 2008
The ATLAS Transition Radiation Tracker (TRT) proportional drift tube: design and performance
\textit{JINST} \textbf{3} P02013

\bibitem{ATLAS} G Aad {\it et al.} The ATLAS Collaboration 2008
The ATLAS Experiment at the CERN Large Hadron Collider.
\textit{JINST} \textbf{3} S08003

\end{thebibliography}
\end{document}